\documentclass[useAMS,usenatbib]{mn2e}

\usepackage{amssymb} % Note

\usepackage{graphicx} % Include figure files

\renewcommand{\H}{{\cal H}}
\newcommand{\WMAP}{{\slshape WMAP~}}

\newcommand{\WMAPc}{{\slshape WMAP}}
\newcommand{\CAMB}{{\sc camb}}

\newcommand{\fsky}{f_{\rm sky}}
\newcommand{\beam}{B_\ell}	% The beam function

\newcommand{\mnras}{{MNRAS}}
\newcommand{\aj}{{AJ}}
\newcommand{\apjl}{{ApJL}}
\newcommand{\prd}{{PhRvD}}
\newcommand{\nat}{{Nat}}
\newcommand{\apj}{{ApJ}}
\newcommand{\prl}{{PhRvL}}

\title[Constraining dark energy using the SKA]{Constraining the Nature of Dark Energy using the SKA}
\author[A.~Torres-Rodr\'iguez \& C.~M.~Cress]
  {A.~Torres-Rodr\'iguez,$^1$ C.~M.~Cress$^1$\\
$^1$Astrophysics and Cosmology Research Unit, University of KwaZulu-Natal, Westville, 4000, South Africa}

\date{\today}

\pagerange{\pageref{firstpage}--\pageref{lastpage}} \pubyear{2006}

\def\LaTeX{L\kern-.36em\raise.3ex\hbox{a}\kern-.15em
    T\kern-.1667em\lower.7ex\hbox{E}\kern-.125emX}

\begin{document}

\label{firstpage}

\maketitle

\begin{abstract}
We investigate the potential of the Square Kilometer Array Telescope (SKA) to constrain the sound speed of dark energy. The Integrated Sachs Wolfe (ISW) effect results in a significant power spectrum signal when CMB temperature anisotropies are cross-correlated with galaxies detectable with the SKA in HI. We consider using this measurement, the autocorrelation of HI galaxies and the CMB temperature power spectrum to derive constraints on the sound speed. We study the contributions to the cross-correlation signal made by galaxies at different redshifts and use redshift tomography to improve the signal-to-noise. We use a $\chi^2$ analysis to estimate the significance of detecting a sound speed different from that expected in quintessence models, finding that there is potential to distinguish very low sound speeds from the quintessence value.
\end{abstract}


\section{\label{sec:Introduction}Introduction}

Current observations of the Cosmic Microwave Background (CMB) anisotropy spectra \citep[hereafter \WMAPc$\,$3]{SpergelWMAP3} combined with large scale structure (LSS) surveys \citep{PercivalLSS,TegmarkLSS} and supernova Type Ia (SNIa) experiments \citep{RiessSupernova,PerlmutterSupernova} have provided strong evidence that we are living in an Universe dominated by an unknown and exotic form of matter which has been called dark energy (DE).

Further evidence for this dark component is provided by the detection of a positive cross-correlation between the CMB temperature anisotropy and several LSS surveys through the Integrated Sachs Wolfe (ISW) effect \citep*{AfshordiISW,NoltaISW,BoughnISW,FosalbaISW,CabreISW}. This effect arises when CMB photons streaming across the dark energy dominated Universe interact with the decaying gravitational potential wells associated with the foreground LSS. This interaction imprints a secondary anisotropy on the primordial CMB spectrum at large scales and it can be used to probe the nature of dark energy.

Current models of dark energy include a cosmological constant ($\Lambda$), with constant energy density filling space homogeneously, and dynamical scalar field models, such as quintessence, whose energy density varies with time and space. Although the current concordance model, involving $\Lambda$ and cold dark matter in an FRW Universe ($\Lambda$CDM), is consistent with all observations, scalar field models cannot be \emph{a priori} excluded. In fact, the small (but non-vanishing) value observed for the dark energy density ($\rho_{\rm DE}$) allows the possibility of studying dynamical models (e.g.~\citealt*{Zlatev,Armendariz,Amendola}).

A common way of describing DE models is through its equation of state parameter, $w = p_{DE} / \rho_{DE}$, where $p_{DE}$ is the pressure and $\rho_{DE}$ is the energy density. The cosmological constant model corresponds specifically to $w = -1$, whereas a value of $w \neq -1$ is associated with dynamical models. The equation of state parameter affects both the evolution of density fluctuations and the background expansion of the Universe.

Another characteristic that defines the nature of a general matter component is its effective sound speed, $c_i$. For a perfect fluid with adiabatic perturbations there is a simple relation between the background and the sound speed: $c_a^2 = \dot p / \dot \rho$. In imperfect fluids, like scalar field models of DE, however, dissipative processes generate entropic perturbations, and the sound speed is given by the more general relation: $c_s^2 = \delta p_{DE} / \delta \rho_{DE}$ \citep{Hucs2}. The sound speed of DE energy thus affects the growth of density perturbations.

Both the equation of state parameter and the sound speed determine the clustering properties of the DE fluid. Single scalar field models, or quintessence, are characterised by having a speed of sound very close to the speed of light, and DE clustering is determined by $w$ alone, occurring only on scales equal to or larger than the horizon size \citep*{Ma99}. For alternative sound speed models, with $c_s^2 < 1$ (in units of the speed of light, $c$), DE clusters on sub-horizon scales which are determined by the value of $c_s$. An example is the scalar field model named k-essence, characterised by having a non-standard form of kinetic energy with an evolving sound speed \citep{Armendariz}.

The ISW effect presents itself as a probe of the properties of DE \citep{CorasPRL} and several groups have provided new ways of constraining the DE parameters using the cross-correlation between the CMB and different LSS surveys (e.g.~\citealt*{Afshordiman,HuScran,BeanDore,PogosianCons,CorasCons}). In this paper we present the prospects of constraining the properties of DE by using an HI-galaxy survey that may be attainable with the future Square Kilometre Array (SKA) radio telescope using the ISW effect in cross-correlation and the HI-galaxy autocorrelation. In particular, we investigate the prospects for detection of a DE speed of sound different to that of quitessence (i.~e., $c_s^2 = 1$).

The outline of the paper is as follows: in \S \ref{sec:theory} we give an overview of the theory of the ISW effect describing the effects of the DE parameters on the ISW--galaxy and galaxy-galaxy power spectra. In \S \ref{sec:SKA} we introduce and discuss the properties of a model SKA HI redshift distribution and the impact on ISW experiments. In section \S \ref{sec:Method} we deal with the numerical implementation of the theory, the signal-to-noise calculations and the forecast for model detection. Finally, the results are presented and discussed in \S \ref{sec:Results}, and they are summarised in \S \ref{sec:Conclusions}.

\section{Theory}\label{sec:theory}

\subsection{\label{sec:Overview}CMB--LSS cross-correlation and galaxy autocorrelation}

Fluctuations in the matter distribution of the large-scale structure can be
expressed in terms of the projected fractional source count of the mass tracer
\begin{equation} 
\frac {\delta N}{N_0}(\hat{n}) = \int_{0}^{z} b_{HI}(z)\,
\frac{d\tilde{N}}{dz}\, \delta_m (z,\hat{n})\, dz \label{galfield}\,,
\end{equation}
where \emph{b$_{HI}$} is the linear bias parameter of an HI-galaxy population, $\delta_m$ is the matter density contrast ($\equiv \delta\rho / \rho$), and $d\tilde{N} / dz$ is the normalised redshift distribution of HI galaxies.

On the other hand, the temperature fluctuations arising from the ISW effect are expressed as the change in gravitational potential over conformal time (or comoving distance)
\begin{equation} 
\frac {\delta T}{T_0}(\hat{n}) = -2 \int_{0}^{r_{dec}} dr\, \dot\Phi (r\hat{n})\,, \label{iswfield}
\end{equation}
where $\Phi$ is the Newtonian gravitational potential, $r$ is the comoving distance
running from today to the epoch of de-coupling and the dot represents derivatives with respect to $r$.

The amplitude of the galaxy--temperature cross-correlation for each mode in Fourier space is thus given
as
\begin{eqnarray}
C_\ell^{gT} &=& 4\pi \int \frac {dk}{k}\; \Big \langle \frac{\delta
N}{N_0}(k) \frac{\delta T}{T_0}(k') \Big \rangle \,j_\ell^2(k r) \nonumber \\
&=& 4\pi  \int_{0}^{\infty} \frac
{dk}{k} f_\ell^N(k)f_\ell^T(k) \Delta_m^2(k)\, , \label{cross}
\end{eqnarray}
where $\Delta_m^2(k)=k^3 P_\delta (k) /2\pi^2$ is the matter power
spectrum per logarithmic interval of wave number ($k$) today and $P_\delta(k)=\langle|\delta(k)|^2\rangle$. The filter
functions $f_\ell^N(k)$ and $f_\ell^T(k)$ in Eq.~(\ref{cross}) correspond to the HI-survey and ISW spectrum respectively. The amount of cross-correlation signal will depend on both the distribution of the galaxy selection function and the change of the gravitational potential with time. This dependency is manifested through the two filter functions which are defined as integrals over redshift weighted by spherical Bessel functions, $j_\ell(k r(z))$, as
\begin{eqnarray}
f_\ell^N(k) &=& \int_0^z b_{HI}(z)\frac{d\tilde{N}}{dz}D(z)\,j_\ell(k r(z))\,dz\,,
\label{filtergal} \\
f_\ell^T(k) &=& \frac{3H_0^2\Omega_m}{c^2 k^2}\int_0^z \frac{dg(z)}{dz}\,j_\ell(k
r(z))\,dz\,,   \label{filterisw}
\end{eqnarray}
where $H_0$ is the value of the Hubble constant today and $\Omega_m$ is the matter density today in units of critical. The function $g(z)$ is the growth suppression factor which is related to the linear growth of matter perturbations
as $g(z)\!=\!(1+z)D(z)$, with $D(z)$ given by $\delta_m(k,z)\!=\!D(z) \delta_m(k,z\!=\!0)$.

Similar to Eq.~(\ref{cross}), the galaxy autocorrelation coefficients (the galaxy angular power spectrum) can be obtained via
\begin{eqnarray}
C_\ell^{gg} &=& 4\pi \int \frac {dk}{k}\; \Big \langle \frac{\delta
N}{N_0}(k) \frac{\delta N}{N_0}(k') \Big \rangle\, j_\ell^2(k r) \nonumber \\
&=& 4\pi  \int_{0}^{\infty} \frac
{dk}{k}\, [f_\ell^N(k)]^2 \Delta_m^2(k)\: . \label{autogal}
\end{eqnarray}

\subsection{The effect of DE cosmological parameters}\label{sec:effect}

The background expansion of the Universe, determined by the Hubble parameter $H(z)$, is affected by the density of the matter components and by $w$. The comoving distance appearing in the previous expressions is defined as
\begin{equation}
r = \int_0^z dz' / H(z')\,, \label{r}
\end{equation}
where, for a flat Universe and in an epoch where radiation is negligible, $H(z)$ is given by
\begin{equation}
\frac{H(z)}{H_0} = \sqrt{\Omega_m(1+z)^3 + \Omega_{DE}\exp\Big[3\int_0^z dz \frac{1 + w(z)}{1 + z}\Big]}\label{H}\,,
\end{equation} 
where we allow for a time-dependent $w$. For a constant equation of state parameter, Eq.~(\ref{H}) simplifies
\begin{equation}
H(z) = H_0\, \sqrt{\Omega_m\,(1+z)^3 + \Omega_{DE}\,(1 + z)^{3(1+w)}}\, . \label{H2}
\end{equation} 

The dark energy term in Eq.~(\ref{H2}) has also an effect on the growth of matter fluctuations. For quintessence models, on scales smaller than the horizon, dark energy can be considered smooth and the matter overdensity grows as (see e.g.~\citealt{Peebles})
\begin{equation}
\ddot\delta_m + 2 H \dot\delta_m - \frac{3}{2} H^2 (1-\Omega_{\Lambda})\delta_m = 0. \label{deltam}
\end{equation}
Eq.~(\ref{deltam}) is usually rewritten as a function of the linear growth factor, $D(z,w)$, and can be solved numerically to compute, for example, the power spectra in the above equations for each $w$ model.

More generally, the introduction of DE perturbations, characterised by $c_s^2$, also has an impact on the evolution of CDM overdensities. This effect is, however, less significant for smaller $w$ and it vanishes as $w\rightarrow-1$ (in fact, it is only in this limit when one could safely neglect DE perturbations). In the synchronous gauge \citep{EvoEqu}, the evolution of the density and velocity perturbation of a general matter component (denoted by subscript \textit{i}) in the frame comoving with the DE (denoted by the circumflex \mbox{$(\,\hat{}\,)$}) is given by \citep{WellerLewis}
\begin{eqnarray}
\delta_i' + 3\H(\hat{c}_{s,i}^2-w_i)(\delta_i +3\H(1+w_i)v_i/k)+ \nonumber \\
(1+w_i)kv_i = -3(1+w_i)h' \label{evoldens}  \\
v_i' + \H(1-3\hat{c}_{s,i}^2)v_i = k \hat{c}_{s,i}^2 \delta_i/(1+w_i) \, ,\label{evolvel}
\end{eqnarray}
where $\H$ is the conformal Hubble parameter, $v_i$ is the velocity and $h'$ is the synchronous metric perturbation \citep{EvoEqu}. As shown in \citet{BeanDore}, the effect of increasing the sound speed of the dark energy is to reduce the growth of structure resulting in an increased ISW signal on the largest scales. On the other hand, as $c_s^2$ decreases, CDM perturbations larger than the DE sound horizon collapse, and the ISW effect is reduced.

Eqs.~(\ref{evoldens}) and (\ref{evolvel}) show how density perturbations are fully characterised by both $w$ and $c_s^2$. The solution to these equations can be obtained numerically, provided that we know these two parameters. In this work, we will consider these to be constants, as justified in \citet{HannestadDE} and references therein, although the formalism can be easily adapted for evolving models of dark energy.

\section{The SKA HI-galaxy redshift survey} \label{sec:SKA}

\subsection{The SKA for ISW experiments}

Direct observation of HI emission from neutral atomic hydrogen\footnote{21 cm line, produced by a change from a parallel to antiparallel spin configuration for the proton and electron.} at high redshift is limited by the poor sensitivity of current radio telescopes. Direct detections have been only achieved up to redshift $z\sim0.18$  \citep*{Zwaan}, and HI gas at higher redshift has only been probed through the absorption in the quasar optical spectra by damped-Ly$\alpha$ (DLA) systems, in which the bulk of the neutral gas in the Universe at high redshift is believed to be contained \citep{Peroux}.

The SKA will have an unprecedented sensitivity provided by its large collecting area, as well as a large ($\sim\!10$ deg$^2$) instantaneous field of view (FOV). The telescope's high sensitivity will allow the detection of HI galaxies located at redshifts that include the beginning of DE domination era, $z\sim$ 0.5\,--\,1, and beyond. This will provide an opportunity to map the sky in different redshifts slices (redshift tomography), and therefore to constrain more accurately the value and evolution of the DE parameters. In addition to a fast surveying capability, an extended FOV will also provide a very wide survey area (almost the entire sky) which is needed to probe the largest scales required by an ISW experiment. All the above will result in the detection of a large number of sources which will reduce statistic errors greatly, and will provide measurements of cross-correlation and galaxy autocorrelation with large signal-to-noise.

\subsection{The HI-galaxy survey}

Our current knowledge of the expected HI galaxy redshift distribution is very limited. A detailed analysis in this regard is presented in \citet[hereafter AR2005]{Abdalla} where the HI mass function, ${\rm d}n/{\rm d}log_{10}M_{\rm HI}$, is integrated over HI mass
\begin{equation}
\frac{{\rm d}N}{{\rm d}z}=\int_{M_{\rm HI}(z)}^{\infty} 
\frac{{\rm d}n}{{\rm d}log_{10}M_{\rm HI}}
\frac{{\rm d}V}{{\rm d}z} {\rm d}log_{10}M_{\rm HI}\,, \label{massfunction}
\end{equation}
where $V$ is the comoving volume and ${M_{\rm HI}(z)}$ is the limiting mass that can be detected by the SKA under certain assumptions. The HI mass function inside this integral is very likely to evolve with redshift due to the fact that the content of neutral Hydrogen is affected by many astrophysical process such as star formation, supernova explosions, etc. The models for the evolution of this HI mass function provided in AR2005 assume the Schechter parametrisation measured by the HIPASS team at redshift $z\sim0$ \citep{Zwaan03}
\begin{equation}
\frac {{\rm d}n}{{\rm d}(M_{\rm HI}/M_{\rm HI}^{\star})}=\theta^{\star} 
\left(\frac{M_{\rm HI}}{M_{\rm HI}^{\star}}\right)^{\alpha} 
\exp\left(-\frac{M_{\rm HI}}{M_{\rm HI}^{\star}}\right) , 
\end{equation}
with $\alpha=-1.3$. The value of the normalisation, $\theta^{\star}$, is set according to the total density of neutral gas $\Omega_{HI}$ found at redshift $z$, which is obtained from DLA observations \citep{Peroux}. The break of the mass function, $M_{\rm HI}^{\star}$, is, however, dependant on the model for galaxy formation and the evolution of the HI mass. In AR2005, three models are presented for this break and we adopt their preferred and more realistic Model `C', which takes star formation into account.

\begin{figure}
\includegraphics[width=9cm,keepaspectratio]{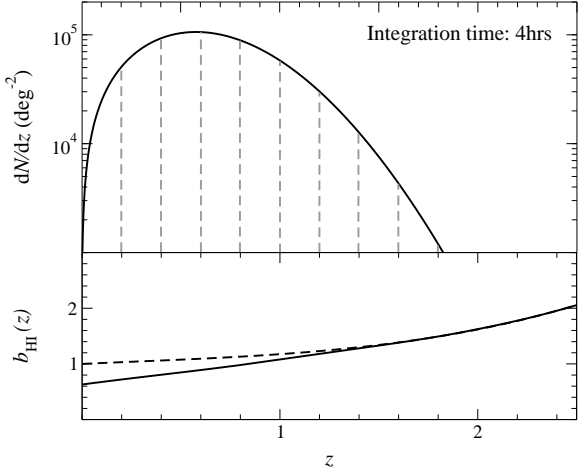}
\caption{Top panel: The number of HI-emitting galaxies per square degree as a function of redshift expected to be detectable with the SKA for an integration time of 4 hours. See text and \citet{Abdalla} for details. Also in grey are represented the redshift slices for $S/N$ measurements (see \S\ref{sec:Method}). Bottom panel: Our model for the bias parameter adapted from \citet{HuScran} (dashed) to match recent HIPASS clustering measurements (solid line) \citep{Basilakos, Meyer}.}\label{dndz}
\end{figure}

The limiting HI mass in Eq.~(\ref{massfunction}) achievable with the SKA is dependant on several observational parameters. For a survey where all redshifts are accessible through single pointing, a FOV frequency dependence of $\nu^{-2}$, a detection threshold of $S/N=10$, and in 4-hour integration time, we use the redshift distribution expressed by the following fitting formula
\begin{equation}
\frac{dN}{dz}={\cal A}\, z \exp\Big(-\frac{(z-z_c)^2}{2\,\Delta z^2}\Big)\,,
\end{equation}
with ${\cal A}=2.52\times10^5$, $z_c=0.211$ and $\Delta z=0.461$. The units are galaxies per square degree. Fig.~\ref{dndz} shows the resulting HI galaxy distribution.

We define our normalised galaxy selection function as $\phi(z) = b_{\rm HI}(z)\times{\rm d}\tilde N/dz$, where the tilde represents the normalised distribution over the redshift of interest. The bias factor used for our analysis (depicted in the lower panel of Fig.~\ref{dndz}) is adapted from the parametrised halo model of \citet{HuJain} (as in \citealt{HuScran}), and modified at low redshift to match recent HIPASS clustering measurements \citep{Basilakos,Meyer}. We varied the bias model and found it has little effect on our estimate for the significance of sound speed detection.

\section{Calculational Method} \label{sec:Method}
\subsection{Numerical implementation}
To test the sensitivity of the cross-correlation to variations in the DE parameters we compute numerically the relevant power spectra (Eqs.~(\ref{cross}) \&~(\ref{autogal}) plus CMB power sectrum) using a modified version of \CAMB\footnote{{\ttfamily http://camb.info.} March 2006 version.} code \citep*{CAMB}. The code takes our galaxy selection function combined with the matter overdensity, $\delta_m(k,z)$, and solves the line-of-sight integral of the resulting ISW source term over conformal time (as in Eq.~(\ref{filtergal})\footnote{Or similar, as \CAMB~computes $\delta_m(k,z)$ from the evolution equations (Eqs.~(\ref{evoldens}) and (\ref{evolvel})) and considers instead the primordial curvature power spectrum in Eq.~(\ref{cross}).}). The output contains the coefficients for the CMB auto-power, $C_\ell^{TT}$, the cross-power with galaxies, $C_\ell^{gT}$ and the galaxy auto-power, $C_\ell^{gg}$.

When considering different models of DE we want to isolate the large-scale variations of the CMB spectrum (i.e., the contribution from the ISW alone). For this reason we fix the angular diameter distance to recombination\footnote{This is well constrained by the position of the first acoustic peak in the \WMAP observations.} and the physical matter density, $\Omega_mh^2$, to be consistent with \WMAPc$\,$3 best fit. To create a family of models, we choose values for $w$ and $h (\equiv H_0 / 100)$ that keep the angular diameter distance fixed, and then vary the DE sound speed, $c_s^2$. To generate the CMB data, we set the optical depth to reionisation, $\tau=0.09$, the primordial power spectrum amplitude, $\Delta_{\mathcal R}^2=2.02\times10^{-9}$, the scalar spectral index, $n_s=0.951$, and the ratio of the initial tensor and scalar power spectrum amplitudes, $T/S=0.55$ (\WMAPc$\,$3).

In order to increase our signal-to-noise, we divide the selection function into several redshift slices (bins) and consider the cross-correlation in each bin as an independent measurement. This assumes that the SKA will be able to measure the redshift of each unresolved galaxy accurately through the HI line. Within this approximation, we can then add up the values of $(S/N)^2$ for each redshift bin in order to obtain the total signal-to-noise of the experiment. The width of the bins is chosen carefully. Wide bins produce larger cross-correlation signal through the ISW filter (Eq.~(\ref{filterisw})), as there is a greater variation of the potential. This, however, increases the total variance as there are fewer measurements along the redshift space. Narrow bins, on the other hand, suffer from a slow-varying gravitational potential and shot noise. Considering the above effects, we choose in our calculations an optimal bin size of $\Delta z = 0.2$ (see Fig.~\ref{dndz}).

\subsection{Signal-to-noise measurement} \label{sec:sn}

The ISW--galaxy cross-correlation signal is affected by various sources of noise such as the cosmic variance, Poisson noise of the galaxy survey, CMB pixel noise and redshift errors (see \citealt{Afshordiman} for more details).

The signal-to-noise ratio squared for a cross-correlation measurement at multipole $\ell$ due to a redshift bin, \emph{i}, is additive
\begin{equation}
(S/N)^2 = \sum_{\ell,\,i} \frac{[C_{\ell}^{gT}\!(w;c_s^2)]^2}{\big [\Delta C_{\ell}^{gT}\!(w;c_s^2)\big ]^2},\label{eq:SN2}
\end{equation}
where the sum extends over redshift bins and multipoles. The expected error $\Delta C_{\ell}^{gT}$ can be obtained from basic principles (e.g.~\citealt{AfshordiISW})
\begin{equation}
\big [\Delta C_{\ell}^{gT}\big]^2 = \frac{1}{(2\ell+1)f^{gT}_{\rm sky}}\big[(C_\ell^{gT})^2 + \tilde C_\ell^{gg}\tilde C_\ell^{TT}\big] \label{variance},
\end{equation}
where $f^{gT}_{\rm sky}$ is the smallest of the CMB and galaxy survey sky coverage fractions and the coefficients $\tilde C_\ell^{gg}$ and $\tilde C_\ell^{TT}$ include the noise power spectra. For the galaxy field, the source of noise comes from the Poisson fluctuations in the number density
\begin{equation}
\tilde C_\ell^{gg}= C_\ell^{gg} + 1 / \tilde n^z_A \,
\end{equation}
where $\tilde n^z_A$ is the galaxy number per steradian in the redshift bin of interest. For the CMB, the noise contribution to the temperature depends on the beam window function and the pixel noise of the experiment (see below).

\subsection{Significance of detection} \label{sec:significance}

In order to determine the significance level of a DE sound speed detection away from the reference model $c_s^2=1$ for a given $w$, we use the $\chi^2$ statistic
\begin{equation} 
-2\log{\cal L}=\chi^2=\sum_{\ell,\,i} \frac{[C_{\ell}^{A}(w;c_s^2)-C_{\ell}^{A}(w;c_s^2=1)]^2}{\big [\Delta C_{\ell}^{A}(w;c_s^2=1)\big ]^2}\,, \label{chi2}
\end{equation} 
where $\mathcal{L}$ is the likelihood of observing the power spectra for a given set of cosmological parameters. In this analysis we assume that, by the time of SKA construction and operation, future independent experiments will provide accurate measurements of parameters other than $c_s$. We will explore potential degeneracies and the possible evolution of $w$ in a follow-up work.

The superscript $A$ in Eq.~(\ref{chi2}) refers to the different observables that constitute our data set, i.e., $C_\ell^{TT}$, $\tilde C_\ell^{gg}$ and $\tilde C_\ell^{gT}$\footnote{In this analysis we ignore CMB polarisation data.}. The errors for the CMB temperature anisotropies and a full-sky ($\fsky^{gal}\!=\!1$) galaxy autocorrelation measurement are respectively
\begin{eqnarray}
\big [\Delta C_{\ell}^{TT}\big]^2 &=&{2\over(2\ell+1)\fsky}(\tilde C_\ell^{TT})^2,\label{TTvariance}\\
\big [\Delta C_{\ell}^{gg}\big]^2 &=& \frac{2}{(2\ell+1)}(\tilde C_\ell^{gg})^2. \label{ggvariance}
\end{eqnarray}

We extend the sum in Eq.~(\ref{chi2}) up to $\ell^{\,gT}_{max}=\ell^{\,gg}_{max}=300$ for the cross-correlation and autocorrelation experiments\footnote{Beyond $l\sim300$, the galaxy autocorrelation becomes dominated by the Poisson term.}, and set $\ell^{\,TT}_{max}=2000$ for the CMB, although the exact value in this case is not critical for sound speed separation through the ISW. Note that we consider independent redshifts measurements and therefore do not include the cross-correlation between bins in Eq.~(\ref{chi2}).

The CMB temperature power spectra, $\tilde C^{TT}_{\ell}\!,$ include the contribution to the pixel noise of the CMB experiment. The observed spectra, i.e.~signal plus noise, are 
\begin{equation} \label{eq:cmbnoise}
\tilde C^{TT}_{\ell} = C^{TT}_{\ell} + w_T^{-1}\beam^{-2}\,,
\end{equation}
where $\beam^2=\exp(-\ell(\ell+1)\theta_{\rm beam}^2/8\ln2)$ is the window function of the (assumed) Gaussian beam, and $\theta_{\rm beam}$ is the beam FWHM. The factor $w_T$ accounts for the detector noise in the temperature measurements. We assume a CMB experiment that measures only temperature anisotropies in two high frequency bands, $143$ and $217$ GHz \citep{Rocha}. In this case, the value of $w_T\beam^{2}$ in Eq.~(\ref{eq:cmbnoise}) is the sum of $w_T\beam^{2}$ for each frequency channel. The parameters for this type of experiment are adapted from \citet{Rocha} and listed in Table~\ref{tab:cmbspecs}.

\begin{table}
\caption{CMB Experimental Specifications. Detector sensitivity $\sigma_T$ is in $\mu$K per FWHM beam and the weight given to one channel is $w_T=(\theta_{\rm beam}\sigma / {\rm T_{CMB}})^{-2}$. For the CMB map we take a sky coverage of $\fsky=0.8$ and thus set $f^{gT}_{\rm sky}=0.8$.}
\label{tab:cmbspecs}
\begin{tabular}{@{}ccc}
\hline
Frequency (GHz) & $\theta_{\rm beam}$ (arcmin) & $\sigma_T$ ($\mu \rm K$) \\
\hline
143 & 8.0 &  6.0  \\
217 & 5.5 &  11.4 \\
\hline
\end{tabular}
\end{table}

Using a similar approach to \citet{PogosianCons}, we combine our three experiments in order to measure the total significance of separation of sound speed models, $(S/N)^2$, using Eq.~(\ref{chi2})
\begin{equation} \label{eq:totalSN}
(S/N)^2_{gT + gal + \rm{CMB}} = {\chi^2}_{gT} + {\chi^2}_{gg} + {\chi^2}_{TT}.
\end{equation}

\section{Results} \label{sec:Results}

%---------------------------------------------------------------------------------------------------

\subsection{The SKA-CMB cross-power spectrum: $C^{gT}$}

\begin{figure}
\begin{center}
\includegraphics[width=8.8cm,keepaspectratio]{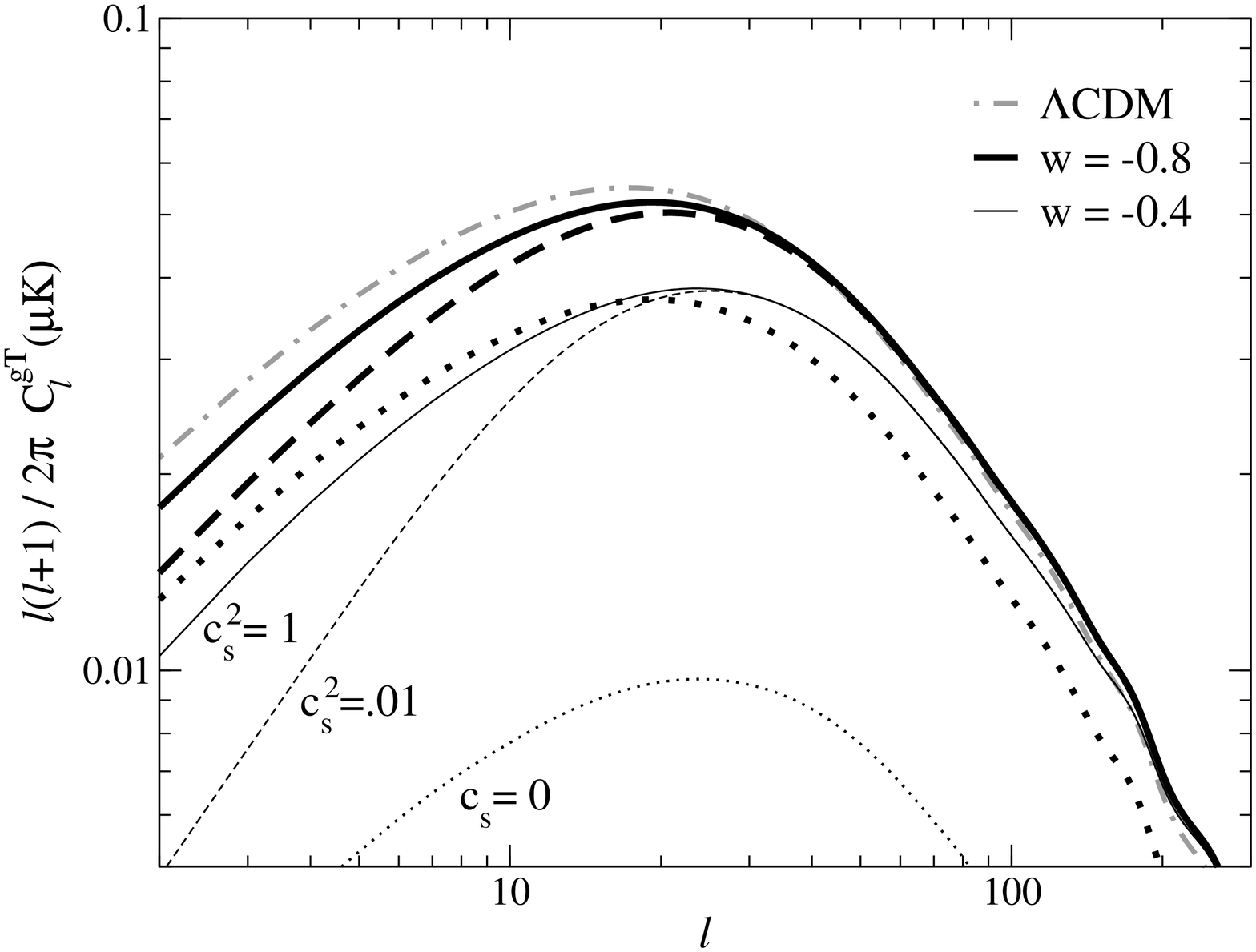}
\caption{Power spectrum from cross-correlation of galaxy overdensities and CMB anisotropies for a single redshift bin $0<z<2$. For the different models, the angular diameter distance to CMB is fixed as well as $\Omega_m h^2$. The cross-correlation becomes more sensitive to the sound speed as $w$ increases.}\label{fig:crosspower}
\end{center}
\end{figure}

Firstly, we show the effect of the sound speed and the equation of state parameter on the cross-correlation spectrum of the SKA HI-galaxy field with the CMB. This is illustrated in Fig.~\ref{fig:crosspower}, where we consider different values of $w$ and $c_s^2$ and a single selection function which covers the redshift range from $z=0$ to $z=2$. Our experiments have shown than this redshift configuration does not provide the optimal $S/N$ which is needed for model comparison. This diagram, however, helps to illustrate the effect of $w$ on the prospects of distinguishing a $c_s^2\neq1$ scenario.

\begin{figure}
\hspace{-.2cm}
\includegraphics[width=9.4cm,keepaspectratio]{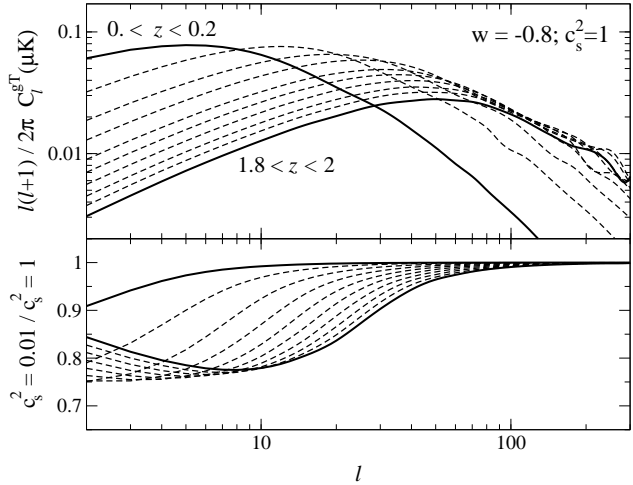}
\caption{Top panel: The cross-power of the binned HI galaxy field with the CMB for a reference model $w=-0.8$ and $c_s^2=1$. Bottom panel: Ratio of the $c_s^2=0.01$ and $c_s^2=1$ spectra. In both panels, the thick solid lines represent the closest and farthest redshift bins, and the intermediate bins are represented by dashed lines.}\label{fig:xsignalbins}
\end{figure}

Fig.~\ref{fig:xsignalbins} shows the cross-power spectrum of each redshift bin referred to in Fig.~\ref{dndz} with the CMB for a reference $w=-0.8$ and quintessence model. As expected, the cross-correlation signal at higher redshifts peaks at smaller angular scales (the galaxy correlation length looks smaller for deeper objects). Because these scales are least affected by cosmic variance, we expect high-redshift objects to contribute greatly in model separation through the cross-correlation. This is seen in the lower panel of the figure where we show the change in the ratio of the spectra for $c_s^2=0.01$ and $c_s^2=1$ models as a function of redshift. The angular scales that provide the largest model separation (the minima in the curves) get smaller (shift to higher $\ell$'s) as we move to higher redshifts.

\subsection{The galaxy autocorrelation: $C^{gg}$}
\begin{figure}
\hspace{-.2cm}
\includegraphics[width=9.4cm,keepaspectratio]{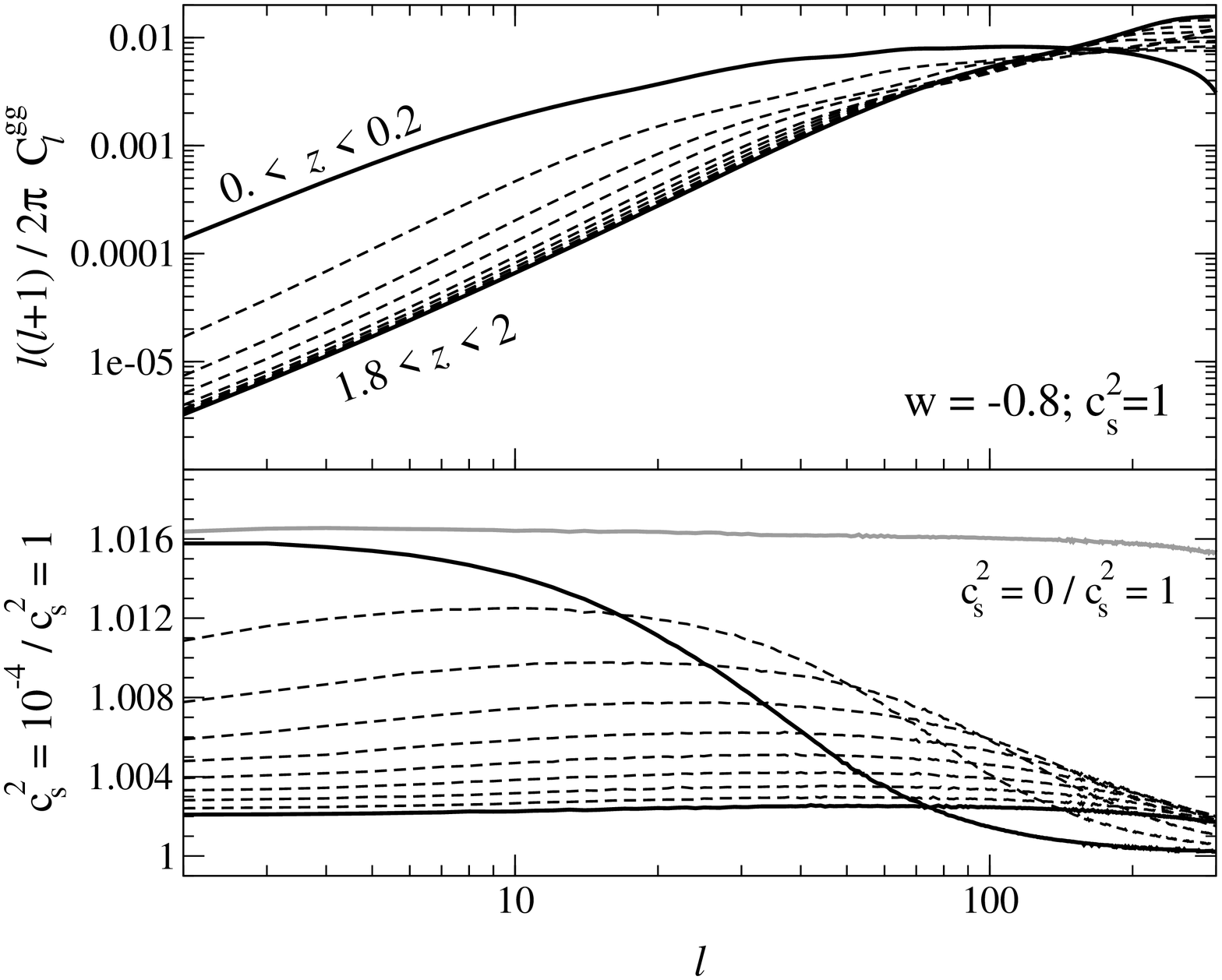}
\caption{Top panel: The angular autocorrelation of the binned HI galaxy field for a reference model $w=-0.8$ and $c_s^2=1$. Bottom panel: Ratio of the $c_s^2=10^{-4}$ and $c_s^2=1$ spectra. Bins are represented as in Fig.~\ref{fig:xsignalbins}. The additional grey line at the top of the panel gives the ratio of the $c_s^2=0$ and $c_s^2=1$ spectra.}\label{fig:galsignalbins}
\end{figure}

We obtain additional information when we look at the angular autocorrelation of the HI galaxies. This is presented in the top panel of Fig.~\ref{fig:galsignalbins} as a function of redshift for the quintessence model. This model is compared with the $c_s^2=10^{-4}$ scenario in the lower panel. On small angular scales, the value of the sound speed has little impact on the galaxy perturbations and we recover the smooth scenario of DE. This effect reduces our ability for model separation using the autocorrelation on the smallest scales, although it is slightly compensated by the shift of the curves towards larger $\ell$'s caused by projection effects (as mentioned earlier). The suppression of the curves' amplitude with redshift, caused by the onset of DE domination occurring only at recent times, affects also our ability for model separation. We include, in addition, the ratio of the $c_s^2=0$ to the smooth case for the lowest redshift bin, and find the discriminating power to be almost scale-independent as, in this limit, DE clusters on all scales. This is an advantage that will become visible when we present the significance for detecting $c_s\rightarrow0$ models.

%------------------------------------------------------------------------------------

\subsection{Cross-correlation signal-to-noise}

\begin{figure}
\includegraphics[width=8.8cm,keepaspectratio]{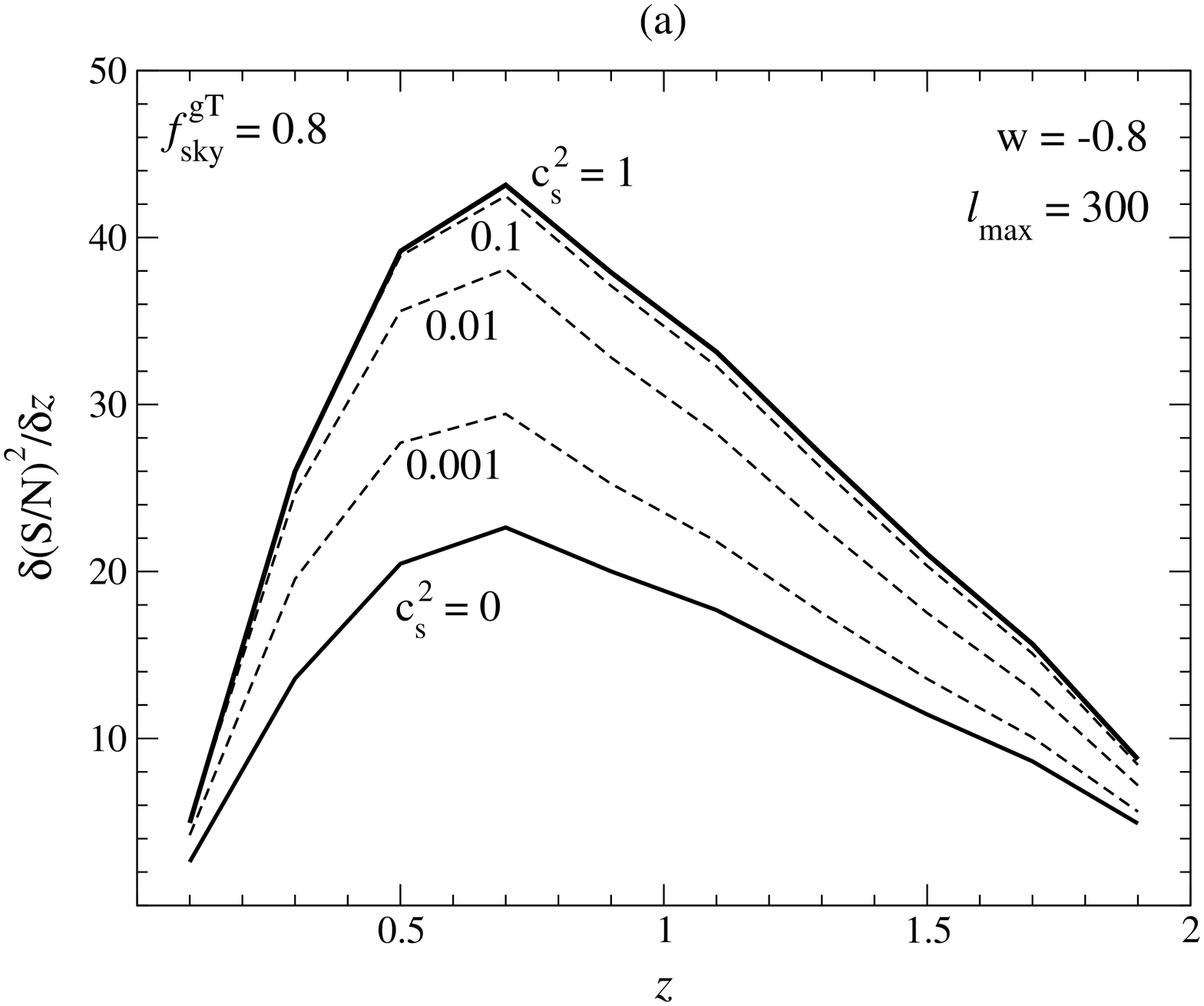}
\includegraphics[width=8.8cm,keepaspectratio]{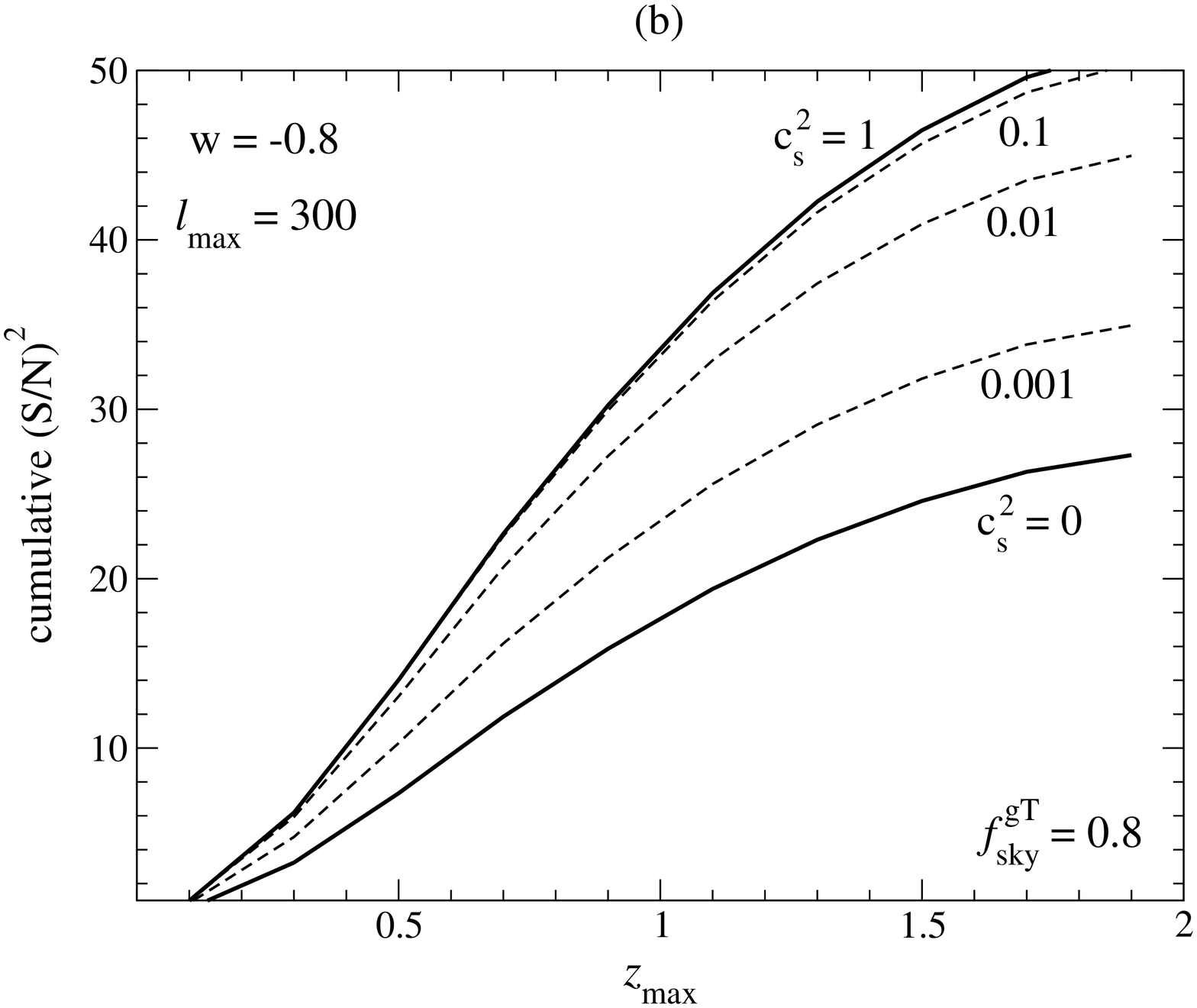}
\caption{(\emph{a}) For the ISW effect in cross-correlation: The differential $(S/N)^2$ per unit redshift for various DE sound speeds as a function of redshift. (\emph{b}) The total $(S/N)^2$, resulting after adding up the contribution of each redshift bin with $z\leq z_{\rm max}$.}\label{SN2}
\end{figure}

Next, we want to investigate the effect of the sound speed and redshift on the cross-correlation signal-to-noise. In Fig.~\ref{SN2} we show the $(S/N)^2$ for a $f^{gT}_{\rm sky}=0.8$ survey in a $w=-0.8$ cosmology and different values of the sound speed as a function of redshift. Fig.~\ref{SN2}a gives the differential $(S/N)^2$ per unit redshift, whereas Fig.~\ref{SN2}b shows the total (summed) $(S/N)^2$ up to redshift $z_{\rm max}$. Although the cross-correlation signal-to-noise peaks at the epochs of DE domination, the results on Fig.~\ref{SN2}b suggests that higher redshifts are required in order to attain a reasonable $(S/N)$ for $c_s^2 \rightarrow0$ models.

%-------------------------------------------------------------------------------------

\subsection{Significance of DE sound speed detection}

\begin{figure}
\includegraphics[width=9.8cm,keepaspectratio]{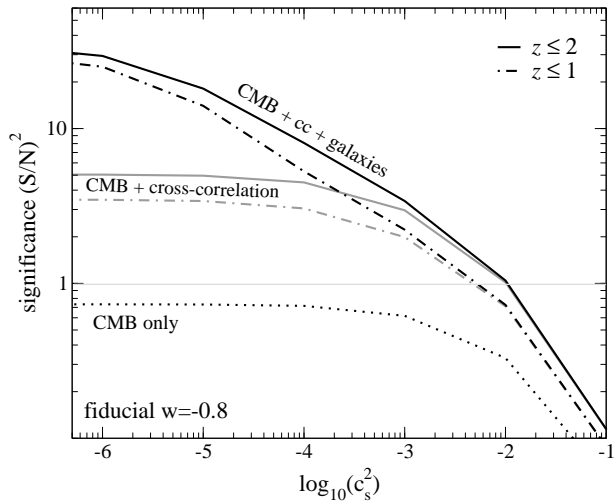}
\caption{The significance of separation between cosmologies with $c_s^2\neq1$ and quintessence with all other parameters fixed and fiducial value $w=-0.8$. Information of the galaxy autocorrelation combined with the ISW--galaxy cross-correlation increases the $S/N$ as $c_s^2\rightarrow0$. Sound speed measurements are improved by the use of higher redshift ($z>1$) objects.}\label{fig:chi2}
\end{figure}

Finally, we present our main results regarding the capability of the SKA to separate different sound speed models of DE. The significance of separation between different models of sound speed with respect to the quintessence model is shown in Fig.~\ref{fig:chi2} for a $w=-0.8$ model and all the other cosmological parameters kept fixed. The significance increases gradually as the sound speed approaches $c_s^2=0$, but falls off rapidly as $c_s^2\rightarrow 1$. In the figure we show the effect of combining the three separate experiments we referred to in \S~\ref{sec:significance}. The CMB experiment gives poor constraints on its own, whereas a detection of $c_s=0.01$ at $S/N=2$ is achieved by cross-correlation using the full redshift survey and at $S/N=3$ when the galaxy information is included. As $c_s^2\rightarrow 0$, the effect of adding the galaxy autocorrelation becomes important, reaching $S/N=5.5$ with all the experiments combined. Although the ISW effect arises mostly at low redshifts, deeper measurements can improve the signal-to-noise in cross-correlation by exploiting the contributions from higher-$\ell$ multipoles (see Fig.~\ref{fig:xsignalbins}).

\begin{figure}
\includegraphics[width=9.8cm,keepaspectratio]{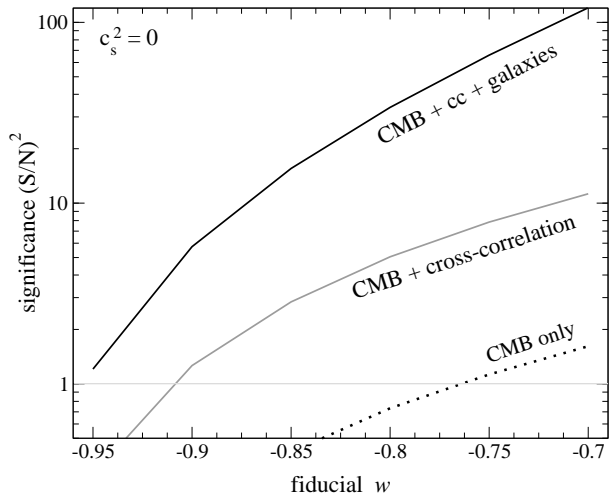}
\caption{The significance of detection of a DE sound speed $c_s^2=0$, for a quintessence model with all other parameters fixed, for a given $w$.}\label{fig:chi2w}
\end{figure}

The significance of detection of a zero sound speed model, $c_s=0$, against a quintessence model is depicted in Fig.~\ref{fig:chi2w} for different given values of the equation of state parameter, $w$. Within the current uncertainty in the value of $w$ (within $10$ per cent of a $\Lambda$CDM model), we find that a detection of $c_s=0$ will have a low significance of $S/N\sim1$ for the $w=-0.95$ case, but this situation improves at $S/N=2.4$, for a $w=-0.9$ Universe. These results give a first estimate of the significance $S/N$ but a full statistical analysis is required before strong conclusions can be drawn. We also note that although current observations do not favour models with $w\gtrsim-0.9$, constraints could be relaxed by considering alternative results (e.g.~\citealt{MacorraNeutrinos}).

\begin{figure}
\includegraphics[width=9.6cm,keepaspectratio]{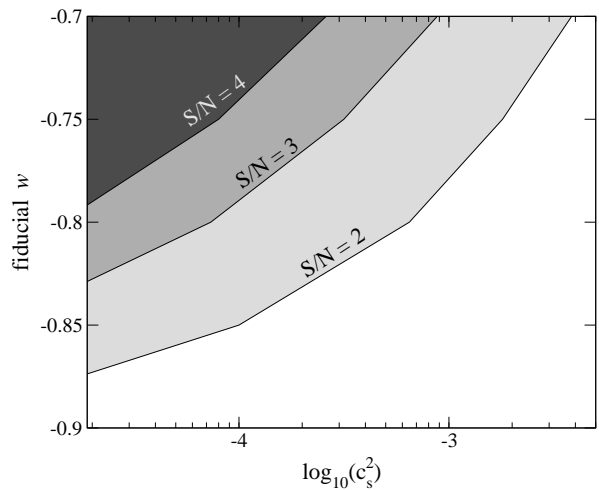}
\caption{Significance boundaries for detection of sound speed for given $w$, in a quintessence model Universe. Models along the boundary lines are detected with the marked significance. The $S/N$ increases as we move along models of increasing $w$ and/or decreasing $c_s^2$. Models with $w\leq-0.9$ can be only detected as $c_s^2\rightarrow0$ (see text).}\label{fig:contours}
\end{figure}

For completeness, Fig.~\ref{fig:contours} summarises our previous results by exploring the significance $S/N$ boundaries for sound speed separation for given values of $w$. As expected (see Fig.~\ref{fig:crosspower}), the significance increases as we move along models with decreasing sound speed for increasing $w$.

\section{Conclusions} \label{sec:Conclusions}

In this paper we have studied the potential of the SKA to constrain different models of dark energy and, in particular, its speed of sound by combining CMB data and an HI-galaxy survey attainable with the SKA. For a constant equation of state parameter, we find that the power spectrum of the cross-correlation is only sensitive to the sound speed as $w$ separates from the cosmological constant limit ($w=-1$). For a reference value of $w=-0.8$, low redshift surveys ($z\sim0.5$) only have the potential for sound speed model separation at the largest scales where cosmic variance dominates, and we find that higher redshift objects ($z\rightarrow2$) are still needed to increase the total signal-to-noise in the cross-correlation. This is particularly important as $c_s\rightarrow0$ where the total cross-correlation $S/N$ is the lowest ($S/N\sim5)$. Our results have not shown a strong dependence on the chosen HI bias model.

In order to forecast the significance of sound speed separation, we have provided a simple $\chi^2$ analysis containing information from three separate measurements: a CMB temperature experiment, the HI-galaxy autocorrelation measurement and the ISW--galaxy cross-correlation. We find that we can only constrain variations of sound speed that are orders of magnitude away from the reference value, $c_s^2=1$. For very small sound speeds, useful constraints can be obtained from the galaxy autocorrelation measurement. As $c_s^2\rightarrow0$, models with $w$ within 10 per cent of $\Lambda$CDM can be separated with at a significance level $2< S/N <3$.

A Fisher matrix analysis including temperature, polarisation and galaxy spectra combined together in a single data covariance matrix will provide more accurate constraints on parameter estimates. A full statistical analysis with marginalisation over other parameters, will be presented elsewhere.

An  HI-galaxy survey with a telescope such as the SKA provides an excellent data set for ISW and power spectrum measurements, both in well-defined redshifts bins. This combination may provide a unique probe of the sound speed of dark energy.

%The ISW effect in cross-correlation with galaxies promises to be a unique probe of the clustering properties of the dark energy. The accurate redshift measurements provided by a SKA HI survey will allow accurate determination of the power spectrum and redshift tomography. Because of this, the SKA promises to be the only experiment that will allow a direct test of the quintessence hypothesis of dark energy.

\section*{Acknowledgements}
The authors would like to thank Kavilan Moodley, Bruce Basset and Robert Lindebaum for helpful discussions and comments throughout the project. ATR acknowledges the government of South Africa through the National Research Foundation (NRF) for their financial support during his PhD.~studies and the SKA project office in South Africa for their support during the course of this project.

\label{lastpage}

%\nocite{*}
%\bibliographystyle{mn2e}
%\bibliography{/Publications/astrodatabase}% Produces the bibliography via BibTeX.

\end{document}